# First application of the failure forecast method to the GPS horizontal displacement data collected in the Campi Flegrei caldera (Italy) in 2011-2020

Andrea Bevilacqua[1], Abani Patra[2], E. Bruce Pitman[3], Marcus Bursik[3], Prospero De Martino[4], Flora Giudicepietro[4], Giovanni Macedonio[4], Stefano Vitale[5], Franco Flandoli[6], Barry Voight[7], Augusto Neri[1].

[1] Istituto Nazionale di Geofisica e Vulcanologia, Sezione di Pisa, Italia; [2] Tufts University, Medford, MA, USA; [3] University at Buffalo, Buffalo, NY, USA; [4] Istituto Nazionale di Geofisica e Vulcanologia, Napoli, Italia; [5] Università di Napoli Federico II, Napoli, Italia; [6] Scuola Normale Superiore, Pisa, Italia; [7] Pennsylvania State University, State College, PA, USA.

**Introduction**

Using the failure forecast method [Voight, 1988] we describe a first assessment of failure time on present-day unrest signals at Campi Flegrei caldera (Italy) based on the horizontal deformation data collected in [2011, 2020] at eleven GPS stations. Campi Flegrei caldera (Italy) is a volcanic field that experienced at least 70 eruptions in the last 15,000 years [Smith et al., 2011; Bevilacqua et al., 2016]. Episodes of slow uplift and subsidence of the ground, called bradyseism, characterize the recent dynamics of the volcanic system [Acocella et al., 2014; Bevilacqua et al., 2020]. In the last decades two major bradyseismic crises occurred, in 1969/1972 and in 1982/1984, with a ground uplift of 1.70 m and 1.85 m, respectively. Thousands of earthquakes, with a maximum magnitude of 4.2, caused the partial evacuation of the town of Pozzuoli in October 1983. They were followed by about 20 years of overall subsidence, about 1 m in total, until 2005 [Del Gaudio et al., 2010]. After 2005 the Campi Flegrei caldera has been rising again with a slower rate than in the previously occurred main phases of uplift, but slowly accelerating [De Martino et al., 2014; Chiodini et al. 2016; Giudicepietro et al. 2019; Tamburello et al. 2019], with a total maximum vertical displacement in the central area of ca. 70 cm (http://www.ov.ingv.it/ov/it/campi-flegrei/monitoraggio.html).

In this study, we apply a probabilistic approach that enhances the well-established method by incorporating a stochastic noise in the linearized equations and a mean-reversion property to constrain it [Bevilacqua et al., 2019]. The stochastic formulation enables the processing of decade-long time windows of data, including the effects of variable dynamics that characterize the unrest of Campi Flegrei caldera. We provide temporal forecasts with uncertainty quantification, giving critical insight into a range of failure times (potentially indicative of eruption dates, see below).

The basis of the failure forecast method is a fundamental law for failing materials: $\dot{w}^{-\alpha} \ddot{w} = A$, where $\dot{w}$ is the rate of the precursor signal, and $\alpha$, $A$ are model parameters that we fit on the data. The solution when $\alpha > 1$ is a power law of exponent $1/(1 - \alpha)$ diverging at time $T_f$, called failure time [Cornelius & Voight, 1995]. In our case study, $T_f$ is the time when the accelerating signals collected at Campi Flegrei would diverge if we extrapolate their trend into the future. The interpretation of $T_f$ as the onset of a volcanic eruption is speculative [Kilburn, 2018].

**Results**

Figure 1 displays the modulus of the GPS horizontal displacement data collected at 11 different stations active from 2011. Three additional stations are not included in the picture because the signal collected is not clearly accelerating; seven additional stations and four GPS buoys were placed after 2011 and will be the target of future analysis. All the stations show an accelerating trend - four stations had a total displacement of ca. 30 cm, five of ca. 20 cm, two of ca. 10 cm. Short episodes of faster displacement are evident in 2012-2013, in 2016 and in 2019-2020. Figure 2 displays the probability forecasts of $T_f$ using the GPS data of 1/2011- 3/2020. The figure is divided in two by a vertical line that represents 3/2020. The left part shows the inverse-rate data collected, and the linearized regression curve that best fits them. The right part shows several examples of stochastic solution paths that extrapolate the data after 3/2020. Moreover, we show the 90% confidence range of $T_f$, and the annual rate of $T_f$, i.e. its probability density function g. The function g is reported as mean values and $95^{th}$ percentile values, due to the uncertainty affecting the parameters of the linearized regression and of the noise properties [Bevilacqua et al., 2019]. The $5^{th}$ percentile values of g are negligible.

**Titolo in italiano: Utilizzo preliminare del failure forecast method sui dati GPS di spostamento orizzontale registrati nella caldera dei Campi Flegrei dal 2011 al 2020**

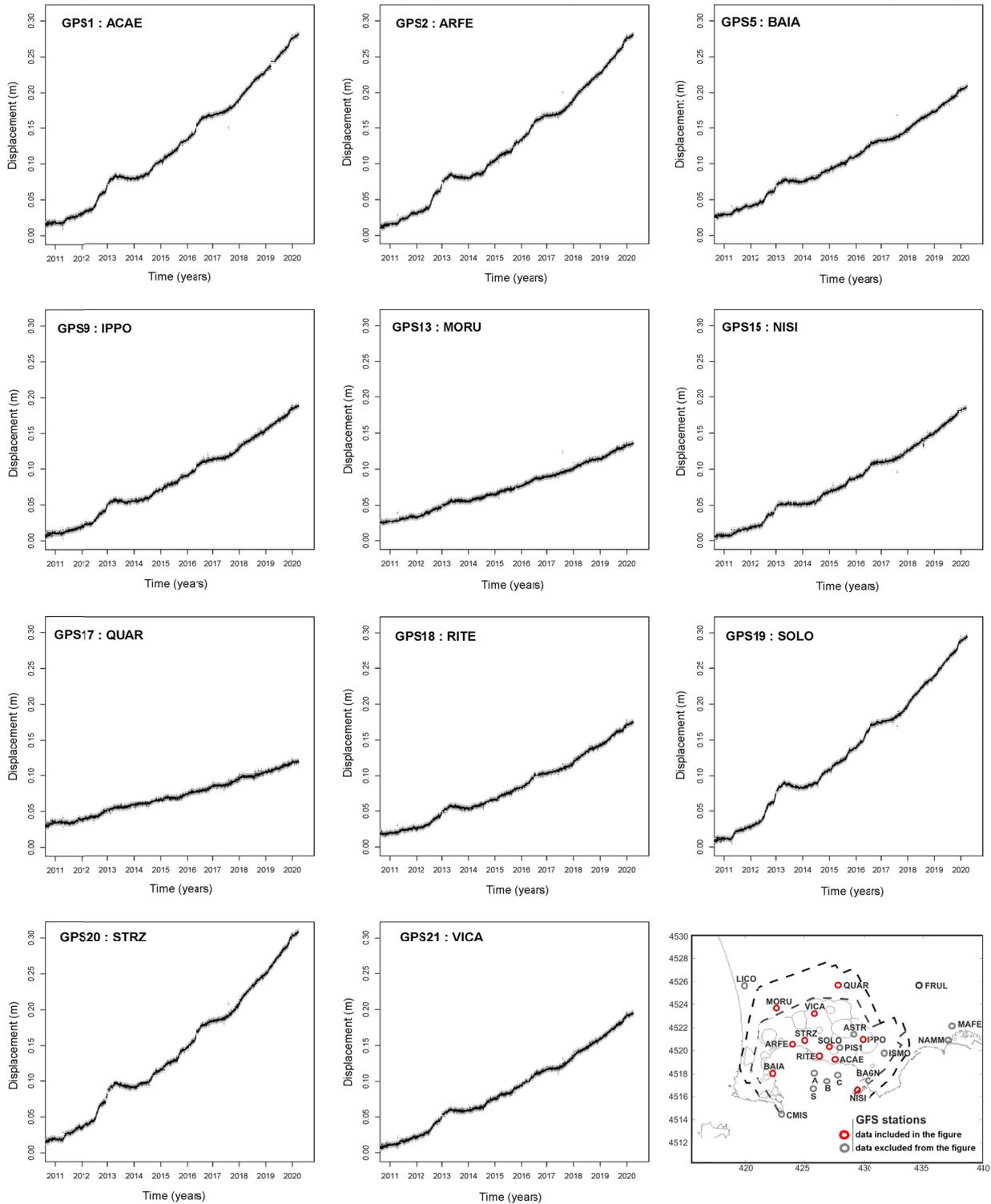

**Figure 1.** GPS horizontal displacement modulus collected in 1/2011-3/2020 at 11 stations mapped in the lower right corner (UTM 33T coordinates). The displacement is computed with respect to the day of deployment, i.e. in year 2000 for GPS 1, 2, 5, 9, 13, 17, 18, in year 2008 for GPS 20, and in year 2009 for GPS 15, 19, 21.

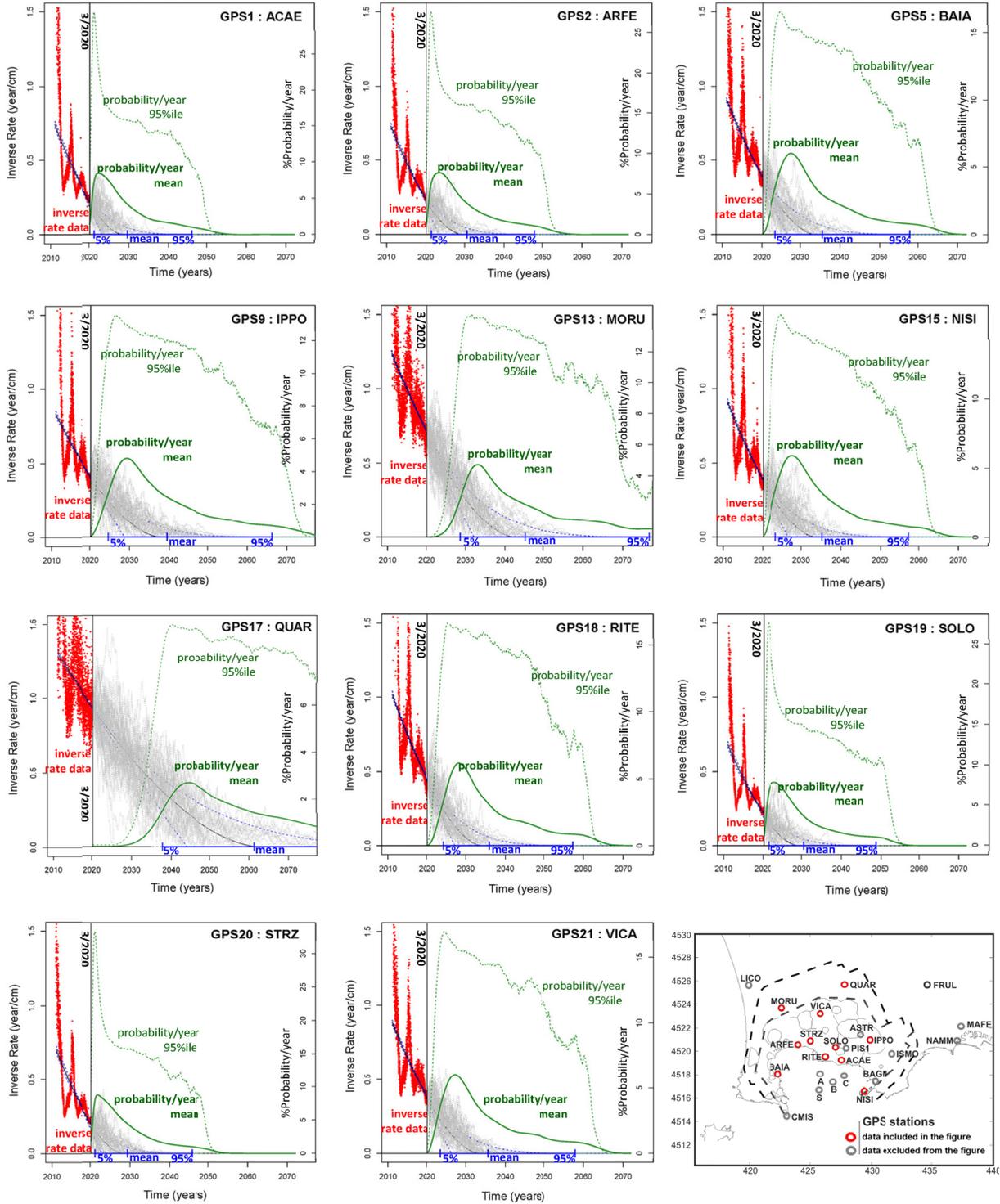

**Figure 2.** Probability forecasts of $T_f$ using the GPS data of 1/2011- 3/2020. Red points on the left are inverse-rate data. The green line on the right is mean values of the annual probability of $T_f$, dashed lines mark its 95th percentile values. A blue line bounds the 90% confidence interval of the forecast. Grey dotted lines display 50 stochastic solution paths. The GPS stations are mapped in the lower right corner (UTM 33T coordinates).

**Discussion and conclusion**

The probability density function g, displayed in Figure 2, is distributed over many decades. The function has peaks of about 12% mean probability per year, and 95$^{th}$ percentile values that can reach 25-30% probability per year. Table 1 shows the probability estimate P that the failure time is realized in 5, 10, or 25 years from 2020. We report the data from the largest to the smallest, showing three groups of estimates. In the first group P is 31-36% in 5 years, 60-64% in 10 years, 92-94% in 25 years. In the second group P is 6.0-12% in 5 years, 28-40% in 10 years, 74-82% in 25 years. In the third group P is 0.0-0.4% in 5 years, 0.2-8.6% in 10 years, 22-63% in 25 years. The three groups correspond to total displacements of ca. 30 cm, 20 cm and 10 cm, respectively.

| Table 1. Campi Flegrei GPS data, horiz. displ. failure time probabilities | | | |
|---|---|---|---|
| GPS station | $P\{T_f < 2025\}$ 5 years | $P\{T_f < 2030\}$ 10 years | $P\{T_f < 2045\}$ 25 years |
| ACAE-1 | 36% | 64% | 94% |
| STRZ-20 | 35% | 62% | 94% |
| SOLO-19 | 32% | 60% | 92% |
| ARFE-2 | 31% | 59% | 92% |
| NISI-15 | 12% | 40% | 81% |
| VICA-21 | 11% | 40% | 82% |
| BAIA-5 | 11% | 39% | 81% |
| RITE-18 | 7.9% | 37% | 82% |
| IPPO-9 | 6.0% | 28% | 74% |
| MORU-13 | 0.37% | 8.6% | 63% |
| QUAR-17 | 0.00% | 0.18% | 22% |

These results provide the starting point for the improvement of short-term hazard assessments that can use monitoring data to adapt the forecast and its uncertainty following a spatio-temporal approach, i.e. short-term vent opening maps [Bevilacqua et al., 2020; Sandri et al, 2020]. It is evident that future variations of monitoring data could either slow down the increase so far observed, or suddenly further increase it leading to shorter failure times than those here reported. Careful spatio-temporal interpolation can provide a full-field view and outlier amelioration.

Although we focus on the ground displacement as an example, a more robust forecasting effort should use multi-sensor data. Several types of unrest signals can be modeled with the failure forecast method, including seismic data and geochemical data [Chiodini et al., 2017; Patra et al., 2019].

**Acknowledgements** In addition to the project "Sale Operative Integrate e Reti di monitoraggio del futuro: l'INGV 2.0", this work is supported by the Dipartimento della Protezione Civile (Italy), as part of the INGV-DPC contract 2019-2021, and by the National Science Foundation award 1821311. The manuscript does not necessarily represent official views and policies of the Dipartimento della Protezione Civile.